\documentclass{llncs}%
\usepackage{makeidx}
\usepackage{url}
\usepackage{graphicx}
\usepackage{amsmath}
\usepackage{amsfonts}
\usepackage{amssymb}%
\begin{document}
%
\title{WRF-Fire Applied in Bulgaria}
\author{Nina Dobrinkova\inst{1} Georgi Jordanov\inst{2} Jan Mandel\inst{3} }
\institute{Institute of Information and Communication Technologies\\
Bulgarian Academy of Sciences,\\
\email{nido@math.bas.bg}
\and Geophysical Institute\\
Bulgarian Academy of Sciences\\
\email{gjordanov@geophys.bas.bg}
\and Department of Mathematical and Statistical Sciences\\
University of Colorado Denver\\
\email{jan.mandel@ucdenver.edu}
}
\maketitle\begin{abstract}
WRF-Fire consists of the WRF (Weather Research and Forecasting Model)
coupled with a fire spread model, based on the level-set method.  We describe a preliminary application of
WRF-Fire to a forest fire in Bulgaria, oportunities for research
of forest fire models for Bulgaria, and plans for the development of  an
Environmental Decision Support Systems which includes computational modeling
of fire behavior.
\end{abstract}%

\section{Introduction}

Forest fires are a problem in most south-European countries, because of the
dry climate during the summer and the year-round high temperatures. Statistics
have been done among the south-European EU member states, where is clear that
the number of the forest fires has increased rapidly in the last 15 years
along with the climate change. In addition, increased pace of development puts
more people and property into harm's way in a wildfire. Bulgaria as part of
this region also has huge problems with wildland fires. Statistics have been
maintained in Bulgaria for the last 30 years, and the number of forest fires
increases every year \cite{Ecopolis-2001,DG-2010}. Even though the number of
wildfires is increasing and the consequences are not only of environmental,
but also of economical and social significance, a proper solution has not been
found neither for Bulgaria nor for the rest of the south EU member states. At
present, most of the countries suffering from wildfires are dealing with the
disaster at the moment of its occurrence. After much controversy, the role of
forest fires as a natural part of the ecosystem and the importance of the fuel
accumulation were recognized in North America, where prescribed burns are now
an integral part of forest management to reduce the fuel, and software tools
can play an important role in the fire management as well in the evaluation of
prescribed burns \cite{Finney-2005-SLE}. Decision support tools integrating
models and observations from a variety of sources are of great interest in
Bulgaria also~\cite{Dobrinkova-2008-PSE}.

Bulgaria has started a nice initiative by opening a center in Sofia on 7th of
July 2007 to provide the disaster control units with adequate, real time
information and to ensure better coordination and effectiveness in the
prevention of natural hazards. It is the first initiative in the policy makers
sector with such orientation. The center is named Aero-Spatial Observation
Center (ASOC). It is aimed to improve and streamline the process of early
warning, prediction and monitoring of natural disasters and accidents on a
national scale. It allows discovering and following the dynamics of wildland
and forest fires, floods, to estimate the loss of forest, control the
conditions of the vegetation, soil humidity and erosion as well as air
pollution. The system operates on national level and is also used by other
governmental organizations e.g. Ministry of Economy and Energy, Ministry of
Environment and Water Supplies, Ministry of Agriculture and Forestry, and
others. The center operates 24/7. However ASOC still lacks a system for
automated satellite image recognition and early detection of forest fires,
floods, and other natural or human-caused disasters. That is why our team from
the Bulgarian Academy of Sciences has started its own research project
dedicated to the forest fires models and tools for fire simulations.

After an analysis of the literature, we have identified WRF-Fire
\cite{Mandel-2009-DAW} as a free Linux-based model which can simulate open
area fires not only in the U.S., but also in Bulgaria, when the run process is
modified to ingest data available in Europe. WRF contains the WRF
Preprocessing System (WPS) \cite[Chapter 3]{Wang-2010-AUG}, which can input
meteorological and land-use data in a number of commonly used formats. WPS has
been extended to process fine-scale land data for use with the fire model,
such as topography and fuel \cite{Beezley-2010-HRW} \cite[Appendix
A]{Wang-2010-AUG}. While the format of meteorological data has largely
stabilized, the ingestion of fire-modeling data was developed for U.S. sources
only, and it may require further preprocessing for other countries.

\section{Coupled Atmosphere-fire Modeling by WRF-Fire}

This section is based on \cite{Mandel-2009-DAW}, where more details can be found.

Fire models range from simple spread formulas to sophisticated computational
fluid dynamics and combustion simulations, see the review in
\cite{Sullivan-2009-RWF}, and also \cite[p. 50]{Mandel-2009-DAW}. However, a
fire behaviour model in a Decision Support System should be faster than real
time in order to deliver a prediction, which dictates a compromise between the
spatial resolution, the processes to be modeled, and fast execution.

Weather has a major influence on wildfire behavior; in particular, the wind
plays a dominant role in the fire progress and shape. Conversely, the fire
influences the weather through the heat and vapor fluxes. Fire heat output can
easily reach the surface intensity of $1\mathrm{MW}/\mathrm{m}^{2}$, and the
fast-rising hot air causes a significant air motion, which affects the
atmosphere also away from the fire. It is known that a large fire
\textquotedblleft creates its own weather.\textquotedblright\ The correct
wildland fire shape and progress result from the two-way interaction between
the fire and the atmosphere~\cite{Clark-2004-DCA,Coen-2005-SBE}.

\subsection{Overview of the Software}

WRF-Fire \cite{Mandel-2009-DAW} combines the Weather Research and Forecasting
Model (WRF) \cite{WRF} with a semi-empirical fire spread model. WRF-Fire got
its start in \cite{Patton-2004-WCA}, where a combination of the tracer-based
model from \cite{Clark-2004-DCA} with WRF was proposed, a~road map was
formulated, and the fundamental observation was made that the innermost
domain, which interacts directly with the fire model, needs to run in the
Large Eddy Simulation (LES) mode. However, instead of tracers, the fire code
in WRF-Fire was developed \cite{Mandel-2009-DAW} based on the level-set method
\cite{Osher-2003-LSM}, partly because the level-set function can be
manipulated more easily than tracers for the purposes of data assimilation.
The code in WRF-Fire for the fire spread rate and feedback to the atmosphere
was taken from \cite{Clark-2004-DCA,Coen-2005-SBE} without any significant
changes, and the initial code for the WRF interface was taken
from~\cite{Patton-2004-WCA}.

In the semi-empirical model, the fire spread rate in the normal direction to
the fireline is assumed to be a function of the fuel properties, the wind
speed close to the ground, and the terrain slope. The fraction of the fuel
left is assumed to be an exponential function of the time from ignition. The
semi-empirical formulas were derived from laboratory experiments, and the
coupled model was verified on several large fires in an earlier
implementation, called CAWFE \cite{Coen-2005-SBE}, with the fire propagation
by tracers and atmospheric modeling by the Clark-Hall weather code. WRF-Fire
takes advantage of this validation and implements a subset of the physical
model from \cite{Clark-2004-DCA,Coen-2005-SBE}: the physical model is the
same, but the fire spread in WRF-Fire is implemented by the level-set method,
and the weather model is replaced by WRF, a supported standard community
weather code. WRF can be run with several nested refined meshes, called
domains in meteorology, which can run different physical models. WRF-Fire
takes advantage of the mature WRF infrastructure for parallel computing and
for data management. An important motivation for the development of the
WRF-Fire software was the ability of WRF to export and import state, thus
facilitating data assimilation (input of additional data while the model is
running), which is essential for fire behaviour prediction from all available
data~\cite{Mandel-2004-NDD}.

\subsection{Mathematical Methods}

Mathematically, the fire model is posed in the horizontal $(x,y)$ plane. The
semi-empirical approach to fire propagation used here assumes that the fire
spreads in the direction normal to the fireline at the speen given by the
modified Rothermel's formula%
\begin{equation}
S=\min\{B_{0},R_{0}+\phi_{W}+\phi_{S}\}, \label{eq:spread}%
\end{equation}
where $B_{0}$ is the backing rate (spread rate against the wind), $R_{0}$ is
the spread rate in the absence of wind, $\phi_{W}=a(\vec{v}\cdot\vec{n})^{b}$
is the wind correction, and $\phi_{S}=d\nabla z\cdot\vec{n}$ is the terrain
correction. Here, $\vec{v}$ is the wind vector, $\nabla z$ is the terrain
gradient vector, and $\vec{n}$ is the normal vector to the fireline in the
direction away from the burning area. In addition, the spread rate is limited
by $S\leq$ $S_{\max}$. Once the fuel is ignited, the amount of the fuel at
location $\left(  x,y\right)  $ is given by%
\begin{equation}
F\left(  x,y,t\right)  =F_{0}(x,y)e^{-({t-t_{i}(x,y)})/T{(x,y)}},\quad
t>t{-t_{i}(x,y)} \label{eq:fuel}%
\end{equation}
where $t$ is the time, $t_{i}$ is the ignition time, $F_{0}$ is the initial
amount of fuel, and $T$ is the time constant of fuel (the time for the fuel to
burn down to $1/e$ of the original quantity). The coefficients $B_{0}$,
$R_{0}$, $a$, $b$, $d$, $S_{\max}$, $F_{0}$, and $T$ in (\ref{eq:spread}) and
(\ref{eq:fuel}) are data.

The heat fluxes from the fire are inserted into the atmospheric model as
forcing terms in the differential equations of the atmospheric model into a
layer above the surface, with exponential decay with altitude. The sensible
heat flux is inserted as the time derivative of the temperature, while the
latent heat flux as the time derivative of water vapor concentration. This
scheme is required because atmospheric models with explicit timestepping, such
as WRF, do not support flux boundary conditions. The heat fluxes from the fire
to the atmosphere are taken proportional to the fuel burning rate, $\partial
F\left(  x,y,t\right)  /\partial t$. The proportionality constants are again
fuel coefficients.

For each point in the plane, the fuel coefficients are given by one of the 13
Anderson categories \cite{Anderson-1982-ADF}. The categories are developed for
the U.S. and different countries use different fuel schemes. WRF-Fire provides
for the definition of the categories as input data, which allows the software
to adapt to other countries.

The burning region at time $t$ is represented level set function $\phi$ by a
as the set of all points $\left(  x,y\right)  $ where $\phi\left(
x,y,t\right)  <0$. It is known that the level set function satisfies the
partial differential equation \cite{Osher-2003-LSM}
\begin{equation}
\partial\phi/\partial t=-S\left\vert \nabla\phi\right\vert ,
\label{eq:level-set}%
\end{equation}
where $\left\vert \nabla\phi\right\vert $ is the Euclidean norm of the
gradient of $\phi$. Equation (\ref{eq:level-set}) is solved numerically by the
finite difference method.

In each time step of the atmospheric model, first the winds are interpolated from
the atmopheric model grid to a finer fire model grid. The numerical scheme for
the level set equation (\ref{eq:level-set}) is then advanced to the next time
step value, the time of ignition is set for any nodes that started burning
during the time step, and the fuel burned during the time step is computed by
quadrature from (\ref{eq:fuel}) in each fire model cell. The resulting heat
fluxes are averaged over the fire cells that make up one atmosphere model
cell, and inserted into the atmospheric model, which then completes its own time step.

\section{Initialization and Computational Results}

WRF-Fire v.3.2 is used for the simulation. The model consists of one domain of
size 4 by 4 km, with horizontal resolution of 50 m for the atmosphere mesh, 80
by 80 grid cells, and with 41 vertical levels from ground surface to 100hPa.
There is no nesting. The domain is located 4 km west from village Zheleznitsa
in the south-east part of Sofia district. This domain is covering the low part
of the forest on Vitosha mountain. The ignition line is located in the center
of the domain, it is 345 m long and the ignition is made at 01 APR 2009,
06:00:02UTC (2 seconds after the start of the simulation). The time step used
in this simulation is 0.5 s. The boundary conditions are specified and are
delivered from the WRF preprocessor WPS. The WRF physics parameterizations
used are \cite[Chapter 5]{Wang-2010-AUG}: Microphysics - Lin et al. scheme
(\texttt{mp\_physics = 2}), Longwave radiation - RRTM scheme
(\texttt{ra\_lw\_physics = 1}), Shortwave Radiation - Dudhia scheme
(\texttt{ra\_sw\_physics = 1}), Surface Layer - MM5 similarity
(\texttt{sf\_sfclay\_physics = 1}), Land Surface - 5-layer thermal diffusion
(\texttt{sf\_surface\_physics = 1}), Planetary Boundary layer - Yonsei
University scheme (\texttt{bl\_pbl\_physics = 1}). Instead of real fuel data,
the fuel used in the fire simulation is based on the altitude
(\texttt{fire\_fuel\_read=1}). The large-scale meteorological background data
has 1 degree horizontal resolution and is obtained from NCEP Global Analysis
Data. The input for land cover and land use data is from the standard data
sources of WRF obtained from USGS with 1km resolution with global coverage
(\url{http://edc2.usgs.gov/glcc/glcc.php}). The terrain input is also from the
standard WRF data sources, USGS with 1 km resolution.

Clearly, the resolution of the land cover and the terrain data is too coarse
for realistic high-resolution studies, but this is the first step of the
testing the WRF-Fire capabilities to work with real data in Bulgaria. For
future experiments, more detailed data will be used. We are in the process of
acquiring databases of high-resolution topography and databases of fuel data,
which specify the type of trees (oak, or a type of conifer). We expect to
specify the custom fuel categories available in WRF-Fire to input the data for
Bulgarian mountains.

\begin{figure}[tbp]
\vspace*{0cm}
\par
\begin{center}
\includegraphics[width=10cm] {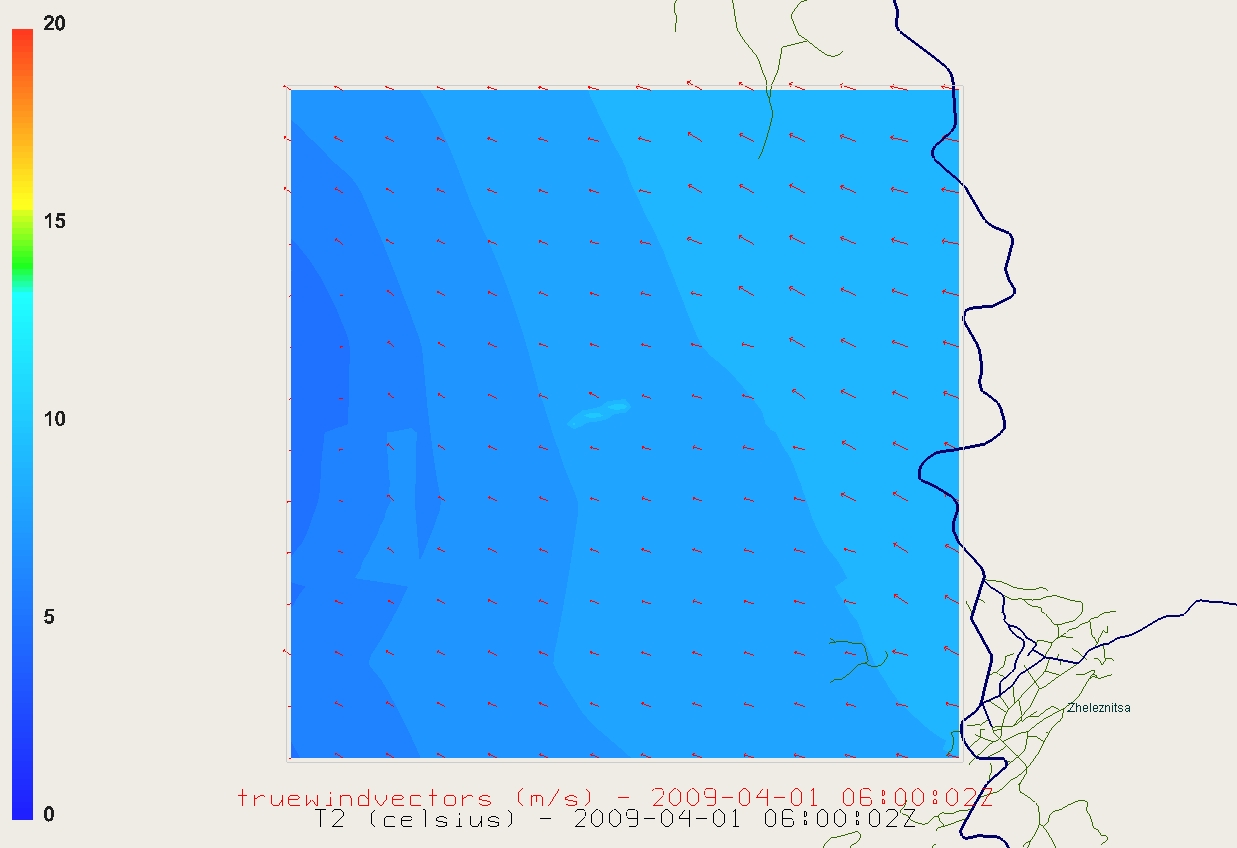}
\end{center}
\par
\caption{Temperature (degrees C) at 2 m above the ground at the time of ignition, 2 s after the simulation start.
The wind vectors are at 10 m height above the ground. The mean 10m wind speed is around 3 m/s. 
The ignition line is visible.}%
\label{T1}%
\end{figure}

\begin{figure}[tbp]
\vspace*{0cm}
\par
\begin{center}
\includegraphics[width=10cm] {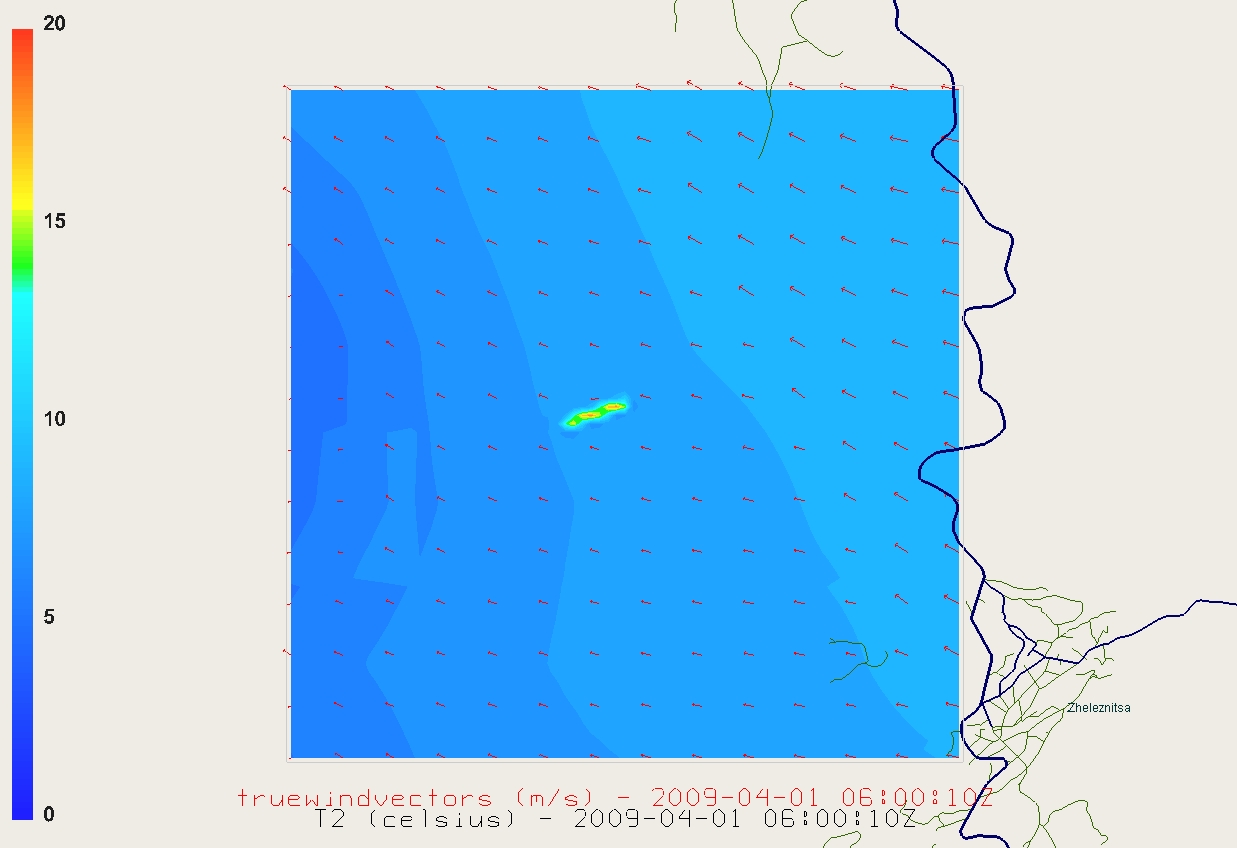}
\end{center}
\par
\caption{The fire is strongly burning and it has spread in the wind direction, 10 s after the simulation start.}%
\label{T3}%
\end{figure}

\begin{figure}[tbp]
\vspace*{0.5cm}
\par
\begin{center}
\includegraphics[width=10cm] {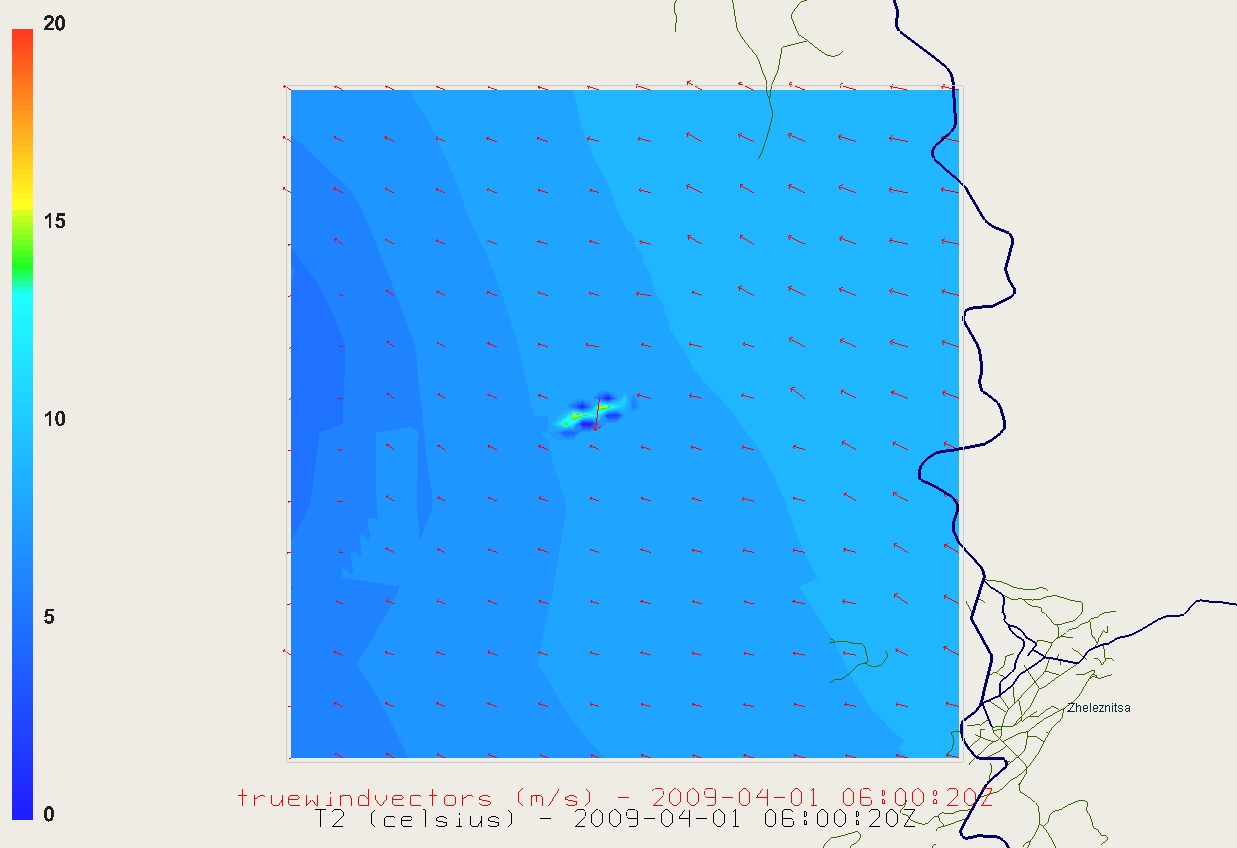}
\end{center}
\par
\caption{The interior of the burning area is cooling down, 20 s after the simulation start.}%
\label{T5}%
\end{figure}

In Figs.~\ref{T1} -- \ref{T5}, we have given the temperature change in the
first stage of the fire propagation, the stage of the fire when the flame is
very strongly burning, and the last picture is with fire intensity slowing
down. The three pictures show also the wind direction and the fire propagation
line. The simulation scenario is a real representation of possible forest fire
at Vitosha mountain, 10 km south from Sofia. The results from this experiment
prove the capabilities of the model to work with real data for areas in
Bulgaria and to give results suitable for forecasting the propagation of the
fire line. The coupling of meteorological model with fire model gives us the
abilities to take into account in our experiments the influence of the wind on
the fire, and, conversely, the winds created by the fire itself. Obtaining as
good meteorological and land data as possible is very important for getting
adequate results in future real cases.

\section{Conclusion and Future Plans}

In this paper we have described how a wildland fire in Vitosha mountain can be
simulated by adapting WRF-Fire v.3.2. Since we were limited by the real data
we had and used idealizations and approximations for the fuel, the fire was
not real, but it does approximate a possible real scenario. We have used raw
data with approximations, because the real data is not available yet for our research.

Our future goal is to incorporate successfuly real data from the wildland
fire, which occured near by the village Leshnikovo in the region of Haskovo,
municipality of Harmanli in August 2009. For the new data, we plan to use 100
meters resolution land-use and land-cover data instead of the 1 km now, and
high-resolution topography along with real fuel data. WRF-Fire is still an
experimental tool for wildland modelling. Our team is aware that most of its
applications are adapted for the U.S. territory and running it with Bulgarian
data has added new features to the model specifics. We intend to make all
results and succesful runs available online for other WRF-Fire users.

\subsection*{Acknowledgements}

This work was supported by the European Social Fund and Bulgarian Ministry of
Education, Youth and Science under Operative Program \textquotedblleft Human
Resources Development,\textquotedblright\ Grant BG051PO001-3.3.04/40, the
project DMU 02-14 \textquotedblleft Collecting and Processing of Data
Concerning Wild land Fires, Occurred on the Bulgarian Territory in the Recent
Years by Using Weather Research and Forecasting Model-Fire
(WRF-Fire)\textquotedblright, and by the U.S. National Science Foundation
under grants AGS-0835579 and CNS-0719641.

\bibliographystyle{splncs03}
\bibliography{NDobrinkova}

\end{document}